\documentclass[conference]{IEEEtran}
\usepackage{cite}
\usepackage{amsmath,amssymb,amsfonts}
\usepackage{algorithmic}
\usepackage{graphicx}
\usepackage{textcomp}
\usepackage{xcolor}

\usepackage{orcidlink} 
\usepackage{hyperref} 
\usepackage{glossaries} 
\usepackage[nameinlink]{cleveref} 
\crefname{figure}{{figure}}{figures}
\Crefname{figure}{{Figure}}{Figures}
\crefformat{footnote}{#2\footnotemark[#1]#3}
\usepackage[scaled=0.9]{helvet} 
\usepackage{flushend} 

\newacronym{jive}{JIVE}{Java Interactive Visualization Environment}
\newacronym{uml}{UML}{Unified Modeling Language}
\newacronym{ide}{IDE}{Integrated Development Environment}
\newacronym{api}{API}{Application Programming Interface}
\newacronym{elk}{ELK}{Eclipse Layout Kernel}
\newacronym{mde}{MDE}{Model-driven engineering}
\newacronym{bpmn}{BPMN}{Business Process Modeling Notation}
\newcommand{\intellij}{IntelliJ IDEA}

\def\BibTeX{{\rm B\kern-.05em{\sc i\kern-.025em b}\kern-.08em
    T\kern-.1667em\lower.7ex\hbox{E}\kern-.125emX}}
\begin{document}

\title{The Visual Debugger Tool\\
{}
\thanks{}
}
\author{
    \IEEEauthorblockN{Tim Kräuter\IEEEauthorrefmark{1}\orcidlink{0000-0003-1795-0611},
    Harald König\IEEEauthorrefmark{2}\IEEEauthorrefmark{1}\orcidlink{0000-0001-6304-6311},
    Adrian Rutle\IEEEauthorrefmark{1}\orcidlink{0000-0002-4158-1644},
    Yngve Lamo\IEEEauthorrefmark{1}\orcidlink{0000-0001-9196-1779}
    }
    \IEEEauthorblockA{\IEEEauthorrefmark{1}Western Norway University of Applied Sciences, Norway}
    \IEEEauthorblockA{\IEEEauthorrefmark{2}FHDW Hannover, Germany
    \\\{tkra, aru, yla\}@hvl.no, harald.koenig@fhdw.de}
}

\maketitle
\begin{abstract}
Debugging is an essential part of software maintenance and evolution since it allows software developers to analyze program execution step by step.
Understanding a program is required to fix potential flaws, alleviate bottlenecks, and implement new desired features.
Thus, software developers spend a large percentage of their time validating and debugging software, resulting in high software maintenance and evolution cost.
We aim to reduce this cost by providing a novel visual debugging tool to software developers to foster program comprehension during debugging.
Our debugging tool visualizes program execution information graphically as an object diagram and is fully integrated into the popular Java development environment \intellij{}.
Moreover, the object diagram allows interactions to explore program execution information in more detail.
A demonstration of our tool is available at \url{https://www.youtube.com/watch?v=lU_OgotweRk}.
\end{abstract}


\section{Introduction}
Debugging is an essential part of software maintenance and evolution since it allows a software developer to analyze program execution step by step.
Nowadays, debugging tools are integrated with every modern \gls*{ide} and are indispensable in software development.
Debugging is used to understand program control- and data flow such that a software developer can locate and fix reported bugs or extend the program to implement new desired features.
Thus, debugging is crucial for software maintenance and evolution, and software developers spend between 35 and 50 percent of their time validating and debugging software \cite{odellDebuggingMindsetUnderstanding2017}.
Consequently, 50-75 percent of the total budget of software development projects is used for debugging, testing, and verification \cite{odellDebuggingMindsetUnderstanding2017}.
We aim to reduce this cost by providing a novel debugging tool to software developers to foster program comprehension during debugging.
Reduced time spent on debugging can be used to implement new features, i.e., create business value for customers.

Traditionally program execution information is represented in a textual manner during debugging (see \cref{fig:variablesIntellij} in \cref{sec:toolDescription}).
Debugging tools integrated with \glspl*{ide}, such as \intellij{} and Eclipse, show the top-level variables contained in the current program scope.
However, the desired program execution information is often not present in the top-level variables but spread out on lower levels of potentially different variables.
Thus, in specific scenarios, a graphical representation results in a faster and better understanding of the shown program execution information.
We have developed a tool that visualizes the current program execution information graphically as an object diagram to foster program comprehension.
This open-source tool, called the \textit{visual debugger}, is integrated with \intellij{}\footnote{The tool is available through the JetBrains Marketplace \cite{VisualDebuggerIntelliJ}.}, which is the most popular Java \gls*{ide} according to the JVM Ecosystem Report 2021 \cite{JVMEcosystemReport2021}.
Compared to other tools, our visual debugger is optimized for industrial use since it is straightforward, lightweight, and non-intrusive.
It can be used alongside the traditional textual debugger and allows interactions to explore program execution information in more detail.
In addition, the tool's architecture enables the reuse of the visualization component in other debugging tools.

The remainder of this paper is structured as follows.
We describe the visual debugger tool in detail (\cref{sec:toolDescription}) and outline a typical usage scenario (\cref{sec:usageScenario}) before explaining the tool architecture (\cref{sec:architecture}).
Finally, we discuss related work in \cref{sec:relatedWork} and conclude in \cref{sec:conclusion}.

\section{Tool description} \label{sec:toolDescription}
\begin{figure}[h]
    \centering
    \includegraphics[width=0.489\textwidth]{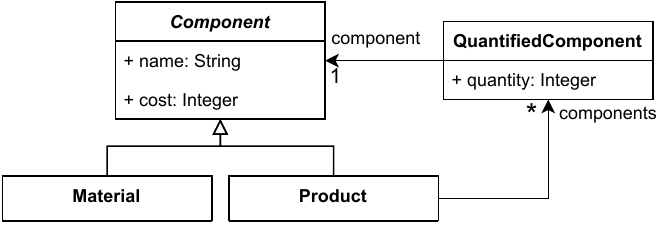}
    \caption{Parts list class diagram}
    \label{fig:partsListModel}
\end{figure}

We will describe our tool using the parts list model shown in \cref{fig:partsListModel}.
A parts list describes the decomposition of \textsf{Products} into sub-products and basic \textsf{Materials}.
Given a parts list, one can calculate the monetary cost and materials needed to construct one or more pieces of a described product\footnote{This example is inspired by the course on information infrastructures taught by Michael Löwe at the University of Applied Sciences FHDW, Hannover.}.

\begin{figure}[h]
    \centering
    \includegraphics[width=0.4\textwidth]{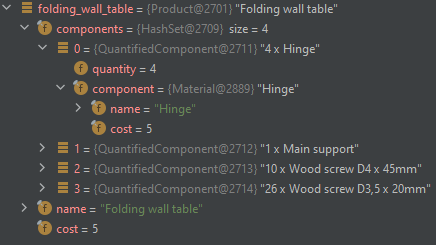}
    \caption{Variables during debugging in \intellij}
    \label{fig:variablesIntellij}
\end{figure}

\Cref{fig:variablesIntellij} shows objects during debugging in \intellij{} conforming to the parts list model. 
The default debugger uses a textual representation for the program execution information.

We have unfolded the substructure of the \textsf{folding wall table} object to see its components, especially the first component, in more detail.
The textual debugging representation is ideal if one is only interested in a small part of the program execution information, such as a single object and its attributes.
However, if the goal is to understand the whole object world, i.e., multiple objects and their links, using the textual representation is not adequate.

Consequently, research on visual debugging began with the goal of fostering program comprehension.
Our tool is one of many visual debugging tools, but we aimed for excellent usability by seamlessly integrating our tool in the debugging process of the \intellij{}.
In addition, our tool is straightforward and non-intrusive, i.e., it complements textual debugging.
The goal of our tool is to make debugging during software development as efficient as possible to increase software developer productivity.

\begin{figure}[h]
    \centering
    \includegraphics[width=0.489\textwidth]{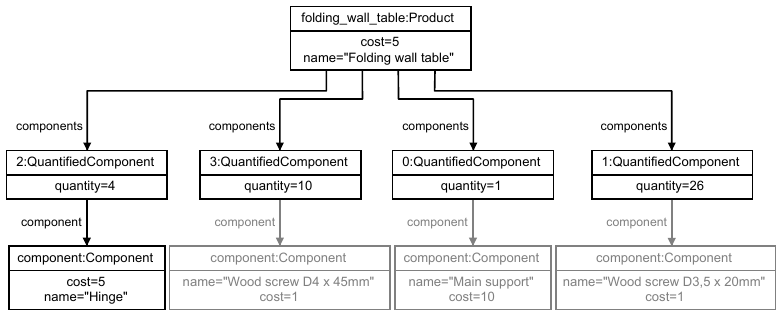}
    \caption{Visual Debugger visualization comparable to \cref{fig:variablesIntellij}}
    \label{fig:visualDebuggerVariables}
\end{figure}

Using our visual debugger tool, we obtain the object diagram shown in \cref{fig:visualDebuggerVariables}\footnote{\label{footnote:artifacts} Additional artifacts, including source code, a demonstration of the visual debugger tool, and a description of the Visual Debugging \acrshort*{api}, can be found in \cite{timkrauterArtifactsICSME2022}.}.
It contains the same objects and level of detail as \cref{fig:variablesIntellij}\footnote{One sees a little more information in \cref{fig:variablesIntellij} due to well-written \textsf{toString()} methods, which are used by \intellij{}.} when ignoring the greyed-out part, which we return to later.

The visual debugger tool continuously visualizes the variables in the scope of the debugging session as an \textit{object diagram}.
The visualization starts automatically when the first breakpoint is reached during debugging if the visual debugger tool is activated.
Thus, if desired, a software developer can use the visual debugger alongside the textual debugging view.
The visualization is always up to date since we listen to the events generated by a debugging session in \intellij{}.
Then, we update the visualization whenever a new breakpoint is reached, or a user steps through the source code.

Textual debugging views only show the top-level variables, i.e., root objects (directly in the debugging session scope) without attributes when debugging is started.
Similarly, we do not visualize all objects linked to the root objects, but we allow the user to configure a \textit{visualization depth}.
The visualization depth describes how many links starting from the root objects should be followed to find objects for the initial visualization.
Afterward, one can explore objects further by double-clicking them in the visualization, just as in the textual debugger.
All objects reachable by outgoing links will be included in the visualization.
For example, in \cref{fig:visualDebuggerVariables}, one quantified component was explored further.
In the future, it can also be interesting to load objects which have outgoing links to the explored object, such that one can load information with and against the link directions.

The visualization is browser-based and implemented in a standalone \textit{visualization component}, which automatically layouts the object diagram using the \gls*{elk}\footnote{\url{https://www.eclipse.org/elk/}}.
The \gls*{elk} layout works well but can be improved to minimize the movement of unchanged objects in the visualization during debugging.
In addition, we provide a visualization based on PlantUML embedded in \intellij{}.
However, it is not possible to explore objects inside the embedded visualization since PlantUML provides static \gls*{uml} diagrams.

The visual debugger tool currently has 2662 unique downloads\footnote{\label{footnote:pluginStats}Last checked on the 21st of Juli, 2022, see \cite{VisualDebuggerIntelliJ}.} and only positive reviews.
It consists of the debugging and the visualization component, which we will describe in more detail in the tool architecture section.
Both components are open-source\cref{footnote:artifacts} and, when combined, result in the visual debugger tool.

\section{Typical usage scenario}  \label{sec:usageScenario}
A typical usage scenario for our tool is debugging a failing unit test.
Unit tests are usually structured according to the \textit{Arrange-Act-Assert} (AAA) pattern.
The \textbf{Arrange} section sets up the unit test context by, for example, initializing a set of needed objects.
Afterward, in the \textbf{Act} section, the method under test is invoked.
Finally, the \textbf{Assert} section verifies that the outcome is as expected.

\begin{figure}[h]
    \centering
    \includegraphics[width=0.488\textwidth]{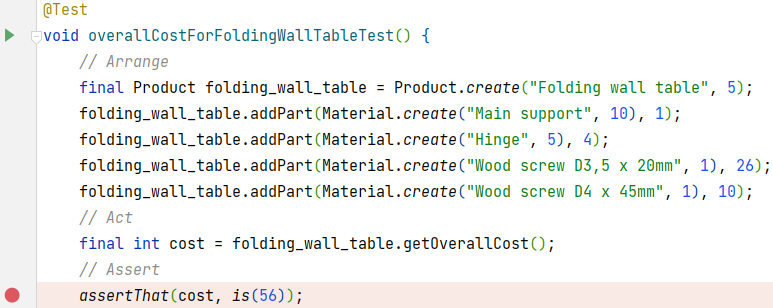}
    \caption{Example java unit test for the parts list in \cref{fig:visualDebuggerVariables}}
    \label{fig:exampleUnitTest}
\end{figure}

\Cref{fig:exampleUnitTest} depicts a \textit{failing} unit test for the parts list model introduced earlier following the AAA Pattern.
In the Arrange section, the objects according to the variables view in \cref{fig:visualDebuggerVariables} \textit{including} the greyed-out part, are created.

In this typical situation, the visual debugger tool can quickly provide an overview of the unit test context created in the Arrange section, i.e., the object world that was set up.
One can set a breakpoint at the start of the Assert section, which will lead to the visualization shown in \cref{fig:visualDebuggerVariables}, including another object containing the computed value from the Act section.
For the example in \cref{fig:exampleUnitTest}, a visualization depth of two or higher is needed.
Otherwise, the first level of objects must be explored one level deep.

The test case is wrong in this example since the expected cost must be increased by 15.
For example, the four hinges in the folding wall table have been counted as one when the expected price was manually computed.
If the test case had been correct, debugging would continue to the invoked method in the Act section.
Obviously, visual debugging can also help understand control and data flow in methods.

Debugging failing unit tests is not our tool's only possible usage scenario since it can be used anytime traditional textual debugging is applicable.
For example, to understand the current codebase and then extend it with new features. 
If desired, one could even use textual and visual debugging simultaneously.

\section{Tool architecture}  \label{sec:architecture}
First, the \textit{debugging component} integrates with \intellij{} by automatically hooking into all started debugging processes of the \gls*{ide}.
The goal of the debugging component is to obtain the current program execution information from \intellij{} and pass it on to the visualization component.
In addition, the debugging component offers a method to load detailed information for individual objects in the current debugging scope, as described earlier.
The debugging component is written in Java, and its code quality and security are continuously checked using static code analysis based on SonarCloud and unit tests \cite{timkrauterArtifactsICSME2022}.

Second, the \textit{visualization component} represents the program execution information as an object diagram to ease program understanding.
Moreover, it allows interaction to load additional program execution information for the currently shown objects.
The visualization component is \emph{browser-based} (JavaScript) and relies on a fixed \emph{Visual Debugging \gls*{api}} \cite{timkrauterArtifactsICSME2022}.
Consequently, we could implement a debugging component for a different \gls*{ide}, such as Eclipse, and reuse the visualization component.
Furthermore, the visualization component is independent of the programming language, which is debugged and can potentially be reused to debug different object-oriented programming languages.

\begin{figure}[h]
    \centering
    \includegraphics[width=0.488\textwidth]{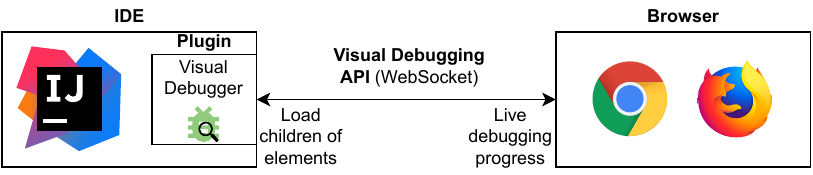}
    \caption{Communication using the Visual Debugger API}
    \label{fig:api}
\end{figure}

The \textit{Visual Debugging \gls*{api}}\cref{footnote:artifacts} is based on \emph{WebSocket} to allow live updates about changes in the program execution information, see \cref{fig:api}.

Initially, a browser connects to the WebSocket server hosting the Visual Debugger \gls*{api}, for example, the server included in our Visual Debugger tool.
Afterward, the browser is updated in real-time about new program execution information due to debugging actions in the \gls*{ide}, such as hitting a breakpoint or jumping to the next line in the source code.
In addition, the visualization component allows a user to interact with the visualization to load all direct children of shown objects.

\begin{figure}[h]
    \centering
    \includegraphics[width=0.488\textwidth]{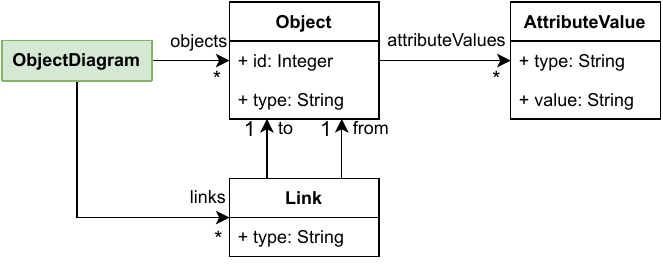}
    \caption{Object diagram metamodel}
    \label{fig:odMetamodel}
\end{figure}

Sending program execution information, i.e., object diagram exchange, is standardized by an XSD schema \cite{timkrauterArtifactsICSME2022}.
\Cref{fig:odMetamodel} depicts the metamodel for object diagrams realized by the schema.

The \textsf{ObjectDiagram} is the root element in the schema (highlighted in green) and contains a set of \textsf{Objects} and \textsf{Links}.
\textsf{Objects} and \textsf{Links} have a \textsf{type}, i.e., the name of a class or association.
In addition, each \textsf{Object} has a unique \textsf{id} provided by the debugger and a set of \textsf{attributeValues}, which have a primitive \textsf{type} and \textsf{value} modeled as strings.

Besides debugging, the visualization component provides two export features.
First, one can export object diagrams during debugging as an SVG file.
This can be useful if an undesired program state has been reached and should be documented in a bug tracking system.
Second, diagrams can be exported as an XML file that can be used to load and edit them in the object diagram modeler, for example, to show the actually desired program state.
The object diagram modeler is an open-source tool to create object diagrams in the browser, developed by the first author \cite{ObjectDiagramModeler2022}.

\section{Related work} \label{sec:relatedWork}
Visual debugging has been researched since the 90s \cite{baeza-yatesVisualDebuggingAutomatic1996, jerdingUsingVisualizationFoster1994, mukherjeaVisualDebuggingIntegrating1994, hansonSimpleExtensibleGraphical1997}, but most of the resulting tools are outdated.
We will now describe recent visual debugging tools and compare them to our tool.

\textit{\gls*{jive}} is a plugin for the Eclipse \gls*{ide} \cite{czyzDeclarativeVisualDebugging2007,k.p.FiniteStateModel2021, JIVEJavaInteractive}.
It provides interactive Java program execution visualization at different levels of granularity.
The program execution information is visualized as a \gls*{uml} object diagram, while the call stack is represented as a \gls*{uml} sequence diagram.
\gls*{jive} is tightly coupled to the Eclipse \gls*{ide} and does not integrate with the Eclipse debugger but rather is a debugging environment on its own.
This approach is significantly different from our tool, which integrates with the debugging tool of the \gls*{ide}.
It makes \gls*{jive} powerful but complex since it is hard to understand what is happening in the multiple views provided by \gls*{jive}.
Compared to \gls*{jive}, the visual debugger tool focuses only on object diagram visualization of the program execution information, making it lightweight and straightforward to use.
In addition, our tool decouples debugging and visualization such that it can be adopted to different \glspl*{ide} even based on other object-oriented programming languages than Java.

A plugin called \textit{Java Visualizer} has been developed for the \intellij{} \cite{JavaVisualizerIntelliJ}.
It visualizes the call stack and objects contained in the Java heap as a box-and-pointer diagram during a debugging session.
However, even in simple scenarios, the visualized call stacks are long since all objects from the Java heap are visualized and not only the variables in the debugging scope.
This leads to much noise in the visualization, especially if one is only interested in the objects currently in the scope of the debugging session.
In contrast, our tool only shows relevant information from the current scope and allows users to load more information if needed.

In \cite{kochGraphicalDebuggingDistributed2015}, the authors describe a tool to debug distributed applications.
It can connect to multiple Java virtual machines and show the retrieved objects separately in an object diagram or combine the same objects from different JVMs using object identifiers or other properties.
The tool is also tightly integrated with the Eclipse \gls*{ide} and tackles the problem of debugging distributed applications, which we do not address.
However, we could not find and test the tool by ourselves.
In the future, we could incorporate these ideas by allowing multiple debugging components (one for each application) to connect to one visualization component.
The visualization component can then show the different debugging views separately or combined as described in \cite{kochGraphicalDebuggingDistributed2015}.

\textit{JAVAVIS} is a standalone tool to help students understand program execution in Java \cite{oechsleJAVAVISAutomaticProgram2002}.
It makes use of object- and sequence diagrams to represent program behavior.
However, it is not integrated with modern \glspl*{ide} such as Eclipse or \intellij{}.
Our tool can help students learn Java or object-oriented program execution in general, but we currently do not provide a sequence diagram visualization.

The Data Display Debugger (DDD) provides a graphical data visualization that can be explored incrementally
and interactively, similar to our approach \cite{zellerDDDFreeGraphical1996}.
However, the tool is not integrated with modern \gls*{ide}s.

Besides source code, \textit{behavioral models} can also be executed and debugged.
For example, \gls*{uml} state-machines, Petri-Nets, or \gls*{bpmn} processes, have clearly defined execution semantics \cite{objectmanagementgroupUnifiedModelingLanguage2017, objectmanagementgroupBusinessProcessModel2013}.
For \gls*{bpmn}, the \textit{bmpn-js token simulation}\footnote{\url{https://bpmn-io.github.io/bpmn-js-token-simulation/}} was developed, enabling the token simulation of \gls*{bpmn} process models in the browser.
The simulator can be seen as a \gls*{bpmn} debugger since one can pause activities, which will stop tokens from flowing through them, similar to breakpoints in source code.
In general, our tool could be adapted to debug behavioral models.
Especially, the visualization component and Visual Debugging \gls*{api} could be extended to visualize behavioral model execution.

\section{Conclusion \& Future work} \label{sec:conclusion}
The main contribution of this paper is the new open-source visual debugging tool, which differs from previously created tools in the following three aspects.
First, it is fully integrated with \intellij{}, a modern and popular \gls*{ide} for Java software development.
According to the JVM Ecosystem Report 2021, over 70\% of JVM developers use \intellij{}  \cite{JVMEcosystemReport2021}.
In addition, the tool received good feedback and was downloaded nearly 2700 times already\cref{footnote:pluginStats}.

Second, the visualization part of the tool is independent, such that it can be reused in other visual debugging tools.
For example, one could develop a plugin for Eclipse \gls*{ide} or Visual Studio Code in the future.

Third, we aimed for the excellent usability of our tool alongside present debugging tools.
Thus, it automatically starts when debugging in \intellij{} and can be used straight away without any configuration.
Moreover, we only show the most relevant program execution information in the debugger by default and allow the user to interactively display more relevant information, similarly to the widely used textual debuggers.

We plan to improve and extend the tool in multiple ways in the future.
First, we want to do more field testing using our tool to gather feedback on its usability and current features.
This should lead to continuous improvement of the tool and greater tool use, which leads to more feedback from practitioners. 
Primarily, the scalability of the tool when debugging large software systems must be investigated.
The scalability of the visual debugger should be similar to the scalability of present textual debuggers, such that our tool is ready for industrial use.
Afterward, we plan a qualitative study to investigate to what extent our tool speeds up software development compared to traditional debugging.

Second, we plan to implement visual debuggers for other \glspl*{ide} and object-oriented programming languages by reusing our visualization component.
The first candidates are Eclipse \gls*{ide} for Java and Visual Studio Code for C\#.

Third, we plan to adapt our tool to debug executions of behavioral models since not only source code can be executed and debugged.
The \textit{bpmn-js token simulation} shows that simulation and debugging benefit software developers using a specific behavioral modeling language.
Furthermore, it is also possible to simultaneously visualize and debug multiple heterogeneous behavioral model executions in heterogeneous modeling situations \cite{krauterBehavioralConsistencyHeterogeneous2021}.
Debugging multiple behavioral model executions simultaneously is similar to debugging distributed applications \cite{kochGraphicalDebuggingDistributed2015}.
\bibliographystyle{IEEEtran}
\bibliography{bib}

\begin{thebibliography}{10}
\providecommand{\url}[1]{#1}
\csname url@samestyle\endcsname
\providecommand{\newblock}{\relax}
\providecommand{\bibinfo}[2]{#2}
\providecommand{\BIBentrySTDinterwordspacing}{\spaceskip=0pt\relax}
\providecommand{\BIBentryALTinterwordstretchfactor}{4}
\providecommand{\BIBentryALTinterwordspacing}{\spaceskip=\fontdimen2\font plus
\BIBentryALTinterwordstretchfactor\fontdimen3\font minus
  \fontdimen4\font\relax}
\providecommand{\BIBforeignlanguage}[2]{{%
\expandafter\ifx\csname l@#1\endcsname\relax
\typeout{** WARNING: IEEEtran.bst: No hyphenation pattern has been}%
\typeout{** loaded for the language `#1'. Using the pattern for}%
\typeout{** the default language instead.}%
\else
\language=\csname l@#1\endcsname
\fi
#2}}
\providecommand{\BIBdecl}{\relax}
\BIBdecl

\bibitem{odellDebuggingMindsetUnderstanding2017}
D.~H. O'Dell, ``The {{Debugging Mindset}}: {{Understanding}} the {{Psychology}}
  of {{Learning Strategies Leads}} to {{Effective Problem-Solving Skills}}.''
  \emph{Queue}, vol.~15, no.~1, pp. 71--90, Feb. 2017.

\bibitem{VisualDebuggerIntelliJ}
``Visual {{Debugger}} - {{IntelliJ IDEs Plugin}} | {{Marketplace}},''
  https://plugins.jetbrains.com/plugin/16851-visual-debugger.

\bibitem{JVMEcosystemReport2021}
``{{JVM Ecosystem Report}} 2021 | {{Snyk}},''
  https://snyk.io/jvm-ecosystem-report-2021/, Jun. 2021.

\bibitem{timkrauterArtifactsICSME2022}
{Tim Kr\"auter}, ``Artifacts - {{ICSME}},''
  https://github.com/timKraeuter/ICSME-2022, Oct. 2022.

\bibitem{ObjectDiagramModeler2022}
``Object diagram modeler,''
  https://github.com/timKraeuter/object-diagram-modeler, Mar. 2022.

\bibitem{baeza-yatesVisualDebuggingAutomatic1996}
R.~A. {Baeza-Yates}, G.~Quezada, and G.~Valmadre, \emph{Visual {{Debugging}}
  and {{Automatic Animation}} of {{C Programs}}}.\hskip 1em plus 0.5em minus
  0.4em\relax {WORLD SCIENTIFIC}, Nov. 1996, vol.~7, pp. 46--58.

\bibitem{jerdingUsingVisualizationFoster1994}
D.~F. Jerding and J.~T. Stasko, ``Using visualization to foster object-oriented
  program understanding,'' {Georgia Institute of Technology}, Tech. Rep., 1994.

\bibitem{mukherjeaVisualDebuggingIntegrating1994}
S.~Mukherjea and J.~T. Stasko, ``Toward visual debugging: Integrating algorithm
  animation capabilities within a source-level debugger,'' \emph{ACM
  Transactions on Computer-Human Interaction}, vol.~1, no.~3, pp. 215--244,
  Sep. 1994.

\bibitem{hansonSimpleExtensibleGraphical1997}
D.~R. Hanson and J.~L. Korn, ``A simple and extensible graphical debugger,'' in
  \emph{Proceedings of the Annual Conference on {{USENIX}} Annual Technical
  Conference}, ser. {{ATEC}} '97.\hskip 1em plus 0.5em minus 0.4em\relax {USA}:
  {USENIX Association}, 1997, p.~13.

\bibitem{czyzDeclarativeVisualDebugging2007}
J.~K. Czyz and B.~Jayaraman, ``Declarative and visual debugging in
  {{Eclipse}},'' in \emph{Proceedings of the 2007 {{OOPSLA}} Workshop on
  Eclipse Technology {{eXchange}} - Eclipse '07}.\hskip 1em plus 0.5em minus
  0.4em\relax {Montreal, Quebec, Canada}: {ACM Press}, 2007, pp. 31--35.

\bibitem{k.p.FiniteStateModel2021}
J.~K.~P., S.~Jayaraman, B.~Jayaraman, and S.~M, ``Finite-state model extraction
  and visualization from {{Java}} program execution,'' \emph{Software: Practice
  and Experience}, vol.~51, no.~2, pp. 409--437, Feb. 2021.

\bibitem{JIVEJavaInteractive}
``{{JIVE}}: {{Java Interactive Visualization Environment}},''
  https://cse.buffalo.edu/jive/.

\bibitem{JavaVisualizerIntelliJ}
``Java {{Visualizer}} - {{IntelliJ IDEs Plugin}} | {{Marketplace}},''
  https://plugins.jetbrains.com/plugin/11512-java-visualizer.

\bibitem{kochGraphicalDebuggingDistributed2015}
A.~Koch and A.~Z{\"u}ndorf, ``Graphical debugging of distributed applications -
  using {{UML}} object diagrams to visualize the state of distributed
  applications at runtime,'' in \emph{Proceedings of the 3rd International
  Conference on Model-Driven Engineering and Software Development}, ser.
  {{MODELSWARD}} 2015.\hskip 1em plus 0.5em minus 0.4em\relax {Setubal, PRT}:
  {SCITEPRESS - Science and Technology Publications, Lda}, 2015, pp. 223--230.

\bibitem{oechsleJAVAVISAutomaticProgram2002}
R.~Oechsle and T.~Schmitt, ``{{JAVAVIS}}: {{Automatic}} program visualization
  with object and sequence diagrams using the java debug interface ({{JDI}}),''
  in \emph{Software Visualization}, S.~Diehl, Ed.\hskip 1em plus 0.5em minus
  0.4em\relax {Berlin, Heidelberg}: {Springer Berlin Heidelberg}, 2002, pp.
  176--190.

\bibitem{zellerDDDFreeGraphical1996}
A.~Zeller and D.~L{\"u}tkehaus, ``{{DDD}}\textemdash a free graphical front-end
  for {{UNIX}} debuggers,'' \emph{SIGPLAN Not.}, vol.~31, no.~1, pp. 22--27,
  Jan. 1996.

\bibitem{objectmanagementgroupUnifiedModelingLanguage2017}
{Object Management Group}, ``Unified {{Modeling Language}}, {{Version}}
  2.5.1,'' https://www.omg.org/spec/UML, Dec. 2017.

\bibitem{objectmanagementgroupBusinessProcessModel2013}
------, ``Business {{Process Model}} and {{Notation}} ({{BPMN}}), {{Version}}
  2.0.2,'' https://www.omg.org/spec/BPMN/, Dec. 2013.

\bibitem{krauterBehavioralConsistencyHeterogeneous2021}
T.~Kr{\"a}uter, ``Towards behavioral consistency in heterogeneous modeling
  scenarios,'' in \emph{2021 {{ACM}}/{{IEEE}} International Conference on Model
  Driven Engineering Languages and Systems Companion ({{MODELS-C}})}, 2021, pp.
  666--671.

\end{thebibliography}

\end{document}